\documentclass[fleqn,12pt,twoside]{article}
\usepackage{espcrc1}

\usepackage{graphicx}
\usepackage[figuresright]{rotating}

\newcommand{\AmS}{{\protect\the\textfont2
  A\kern-.1667em\lower.5ex\hbox{M}\kern-.125emS}}

\title{Temporal variation of coupling constants and nucleosynthesis}

\author{H.~Oberhummer\address{Atominstitut of the Austrian Universities, 
                Vienna University of Technology,\\
                Wiedner Hauptstr.~8--10, 1040 Vienna, Austria}, 
		A.~Cs\'ot\'o\address{Department of Atomic Physics,
E\"otv\"os University,\\
		P\'azm\'any P\'eter s\'et\'any 1/A, 1117 Budapest, Hungary},
		M.~Fairbairn\address{Service de Physique Th\'eorique,
Universit\'e Libre de Bruxelles,\\
		CP225, Bvd.~de Triomphe, 1050 Bruxelles, Belgium},
		H.~Schlattl\address{Astrophysics Research Institute,
Liverpool John Moores University,\\
		Twelve Quays House, Egerton Wharf, Birkenhead CH41 1LD,
United Kingdom} and
		M.M.~Sharma\address{Physics Department, Kuwait University,
Kuwait 13060}
		}

\begin{document}

\maketitle

\begin{abstract}
We investigate the triple-alpha process and the Oklo
phenomenon to obtain constraints on possible cosmological
time variations of fundamental constants.
Specifically we study cosmological temporal constraints
for the fine structure constant and nucleon and meson masses.
\end{abstract}

\section{Introduction}

In previous works, different constraints have been obtained for possible
temporal variations of fundamental constants.
Recent investigations involved
primordial nucleosynthesis
\cite{Kolb:1985sj,Dixit:1987at,Bergstrom:1999,Avelino:2001,Nollet:2002,Ichikawa:2002bt,Dent:2001ga,Flambaum:2002de},
cosmic microwave background (CMB) \cite{Avelino:2001,Martins:2002}, quasar
absorption lines \cite{Webb:2000mn,Murphy:2001}, stellar nucleosynthesis
\cite{Ricci:2002},
meteorites \cite{Ricci:2002,Lindner:1986,Olive:2002tz}, the Oklo
phenomenon
\cite{Flambaum:2002de,Olive:2002tz,Damour:1996,Fujii:2000,Fujii:2002}
and atomic clocks \cite{Flambaum:2002de,Prestage:1995,Sortais:2001}.
In most of these works temporal variations of the fine structure constant
were investigated.
Presently the only analysis showing
a time variation definitely different from zero stems from the
analysis of atomic multiplet spectra in quasar absorption lines through
intervening interstellar clouds
at $0.5 < z < 3.5$ \cite{Webb:2000mn,Murphy:2001}.

In this work we investigate the triple-alpha process and the Oklo
phenomenon to obtain cosmological constraints on the time variation of the
electromagnetic fine structure constant, and the strength of the strong
interaction, i.e., the nucleon and meson masses
or the QCD scale parameter $\Lambda_{\rm QCD}$.

\section{Triple--Alpha process}%

In stellar nucleosynthesis one of the most sensitive reactions
with respect to possible variations of the
coupling constants is the triple--alpha process leading
to the production of $^{12}$C \cite{obe00,obe01}. Stellar nucleosynthesis
of carbon
and oxygen in helium burning of ancient stars
is therefore particularly of interest for two reasons: The process is very
sensitive to variations in fundamental
parameters and also took place at the same time ($t_{\rm B}
\approx 5-13$\,Gyr
or $z \approx 0.5-3$) that absorption lines were being created in the light
from quasars due to intervening gas
clouds \cite{Webb:2000mn,Murphy:2001}.  The C/O abundance ratio and the
absorption lines in the quasar
spectra are therefore both sensitive to variations in coupling constants at
the same epoch.

We compare the change of abundance
ratios of C/O by variation of fundamental parameters like the fine
structure constant $\alpha_{\rm em}$,
and/or the nucleon $m_{\rm N}$ and meson masses $m_{\rm B}$. In the
first step we investigated the changes of
the 0$_2^+$--resonance in $^{12}$C using nuclear microscopic
model calculations \cite{obe00,obe01} (Table 1).

\begin{table}
\begin{center}
\label{tt1}
\caption{Change of the resonance energy $\Delta E_{\rm R}$ in keV of the
0$_2^+$--resonance
in $^{12}$C for variations of the fine structure constant $\alpha_{\rm em}$
in the Coulomb potential,
the nucleon mass $m_{\rm N}$ in the kinetic energy term and the meson
masses $m_{\rm B}$
in the effective nucleon--nucleon potential with a corresponding
enhancement/reduction factor $p$.}
\newcommand{\m}{\hphantom{$-$}}
\newcommand{\cc}[1]{\multicolumn{1}{c}{#1}}
\renewcommand{\tabcolsep}{2pc} 
\renewcommand{\arraystretch}{1.2} 
\begin{tabular}{ccc}
\hline
Variation & \multicolumn{2}{c}{Energy shift (keV)}\\
of parameter(s) & $p=1.001$ & $p=0.999$\\
\hline
$\alpha_{\rm em}$ & $+3.88$ & $-3.89$ \\
$m_{\rm N}$ & $-22.63$  & $+22.44$\\
$m_{\rm B}$ & $+25.87$ & $-26.13$ \\
$m_{\rm N}$, $m_{\rm B}$$^{\rm a}$ & $+3.51$ & $-3.44$ \\
$\alpha_{\rm em}$, $m_{\rm N}$ and $m_{\rm B}$$^{\rm b}$ &
$+(3.59,3.64)$ & $-(3.51,3.58)$ \\
\hline
\end{tabular}
\end{center}
{\footnotesize $^{\rm a}$ We assume that the nucleon mass and the
exchanged meson masses scale identically.}\\
{\footnotesize $^{\rm b}$ We assume that the nucleon mass and the
exchanged meson masses scale identically, whereas the fine structure constant
scales with a factor 30--60 less
\cite{Dent:2001ga,Langacker:2001td,Calmet:2001nu}.}\\
\end{table}

The observation of the C/O--abundance ratios in ancient stars and their
computation
in stellar models do not seem to vary by more than a factor 3 (see e.g.,
\cite{Carigi:2002}).
Such a change would be produced in helium burning of Red Giants by a
variation of
the fine structure constant of not more than $\pm 0.6\,\%$. This leads
to the constraint $\Delta\alpha_{\rm em}/\alpha_{\rm em} \approx \pm 6
\times 10^{-3}$
shown in Table 2. The time variation of the fine structure constant
through the
analysis of atomic multiplet spectra in quasar absorption lines through
intervening interstellar clouds of the order of $\Delta\alpha_{\rm
em}$/$\alpha_{\rm em} \approx
10^{-5}$ is still much weaker than the constraint obtained from the
triple--alpha process
of the order of $10^{-3}-10^{-2}$ (see Table 2).

Almost the same value for the constraint is obtained when assuming identical
linear scaling of the nucleon $m_{\rm N}$ and meson masses $m_{\rm B}$, i.e.,
$\Delta m_{\rm N}/m_{\rm N} =
\Delta m_{\rm B}/m_{\rm B} =
\Delta\Lambda_{\rm QCD}/\Lambda_{\rm QCD}$. That
the same value is obtained by varying either the fine-structure constant or
simultaneously
the nucleon and meson masses (the first and fourth line in Table 1) can
also be verified in first--order perturbation
theory.
In this case, the corresponding constraint of the fine structure constant
is obtained by assuming that changes
in $\alpha_{\rm em}$ are of the order $30-60$ larger than for $\Lambda_{\rm
QCD}$
\cite{Dent:2001ga,Langacker:2001td,Calmet:2001nu}.
Therefore, the constraint obtained for $\alpha_{\rm em}$ is one to two
orders of magnitude more stringent than the one by
varying solely the fine structure constant.

\section{The Oklo phenomenon}

\begin{table}
\begin{center}
\label{tt2}
\caption{Approximate constraints upon time
variations of the fine structure
constant $\Delta\alpha_{\rm em}/\alpha_{\rm em}$ and $\dot{\alpha}_{\rm
em}$/$\alpha_{\rm em}$ (yr)$^{-1}$ for mean look--back times
$\bar{t}_{\rm B}$
or corresponding mean red shifts $\bar{z}$, respectively.}
\newcommand{\m}{\hphantom{$-$}}
\newcommand{\cc}[1]{\multicolumn{1}{c}{#1}}
\renewcommand{\tabcolsep}{2pc} 
\renewcommand{\arraystretch}{1.2} 
\begin{tabular}{cc}
\hline
$\Delta\alpha_{\rm em}/\alpha_{\rm em}$  & $\dot{\alpha}_{\rm
em}$/$\alpha_{\rm em}$ (yr)$^{-1}$ \\
\hline
\multicolumn{2}{c}{Quasar absorption lines \cite{Webb:2000mn,Murphy:2001}:
$\bar{t}_{\rm B} \approx 10$\,Gyr,
$\bar{z} \approx 1.5$}\\
$(-0.72 \pm 0.18) \times 10^{-5}$ & $(-0.72 \pm
0.18) \times 10^{-15}$ \\
\hline
\multicolumn{2}{c}{Triple--alpha process (this work): $\bar{t}_{\rm B}
\approx 10$\,Gyr,
$\bar{z} \approx 1.5$}\\
$\pm 6 \times 10^{-3}$ & $\pm 6 \times 10^{-13}$ \\
\hline
\multicolumn{2}{c}{Oklo phenomenon (\cite{Fujii:2000,Fujii:2002} \& this
work): $\bar{t}_{\rm B} \approx 2$\,Gyr, $\bar{z} \approx
0.1$}\\
$(-3.6 \pm 14.4) \times 10^{-9}$ & $(-1.8 \pm 7.7) \times 10^{-18}$ \\
\hline
\end{tabular}
\end{center}
\end{table}

The Oklo phenomenon that occurred about 2\,Gyr ago gives the most stringent
constraints
for variations of fundamental parameters like the fine structure constant
as it is seen from
Table 2. The Oklo phenomenon refers to a natural
fission reactor which was operating at the Oklo uranium mine in Gabon. By
examining
the isotopic ratios of Sm in the Oklo reactor limits on cosmological time
variations of fundamental parameters can be obtained
\cite{Flambaum:2002de,Olive:2002tz,Damour:1996,Fujii:2000,Fujii:2002}. These
bounds were derived by calculating the energy shift
of a resonant state in $^{150}$Sm lying just 0.0937\,eV above the threshold
of the reaction
$^{149}{\rm Sm} + {\rm n} \rightarrow ^{150}{\rm Sm} + \gamma$.
The quantity of interest is the energy of that state
$E_{\rm R} = Q - E^* = 0.0937$\,eV,
where $Q$ is the Q--value of the reaction and $E^*$ is the excitation energy of
the compound nucleus $^{150}$Sm.

We employ the Relativistic Mean--Field theory (RMFT) \cite{Serot:1992} to
calculate
the ground--state properties of $^{149}$Sm and $^{150}$Sm nuclei.
We obtain the variation in the Coulomb energy
difference between the two nuclei by varying the fine structure
constant  $\alpha_{\rm em}$ fractionally by 10$^{-5}$,
10$^{-4}$ and 10$^{-3}$. The change in the Coulomb energy difference
shows a linear relationship with the variation in $\alpha_{\rm em}$.
We obtain for the Coulomb energy difference between
$^{150}$Sm and $^{149}$Sm a value of 1.18\,MeV, which is only slightly
larger than the values
obtained from the phenomenological Bethe--Weizs\"acker formula. Therefore,
the results claimed in recent work \cite{Damour:1996,Fujii:2000,Fujii:2002}
remain unchanged in the light of RMFT--calculations.

We also varied the masses of the nucleon and the $\sigma$--, $\omega$-- and
$\rho$--mesons. Fractional changes were made of 10$^{-5}$,  10$^{-4}$ and
10$^{-3}$ in the masses. The total energy difference between
the isotopes $^{150}$Sm and $^{149}$Sm shows again a linear behaviour
with the variation in the masses. The change in the total
energy difference turns out to be about a factor of 10 larger
than the above corresponding change in the Coulomb energy due to
the variation in the fine structure constant. This would correspond
to a constraint for the fine structure constant that is again as before at
least
two factors of magnitude more stringent than the one obtained by solely
varying the fine structure constant.

In any case the Oklo phenomenon gives a more stringent constraint
than the analysis of atomic multiplet spectra in quasar absorption lines
as can be seen from Table 2. A possible explanation could be
a non--linear temporal or even a non--monotonic variation of the fine
structure constant as has
been suggested by some recent papers, since the
Oklo phenomenon occurred at a much later time than the absorption lines in
the quasar spectra were created.

\section*{Acknowledgments}
Supported by OTKA-D32513/T037548, FKFP-0242-2000/0147-2001, and by
the John Templeton Foundation (983-COS153).

\end{document}